\documentclass[aps,preprint,amsmath,amssymb]{revtex4}

\usepackage{graphicx}
\usepackage{url}
\usepackage{color}

\usepackage[ pdftex, plainpages = false, pdfpagelabels,
         pdfpagelayout = useoutlines,
         bookmarks,
         bookmarksopen = true,
         bookmarksnumbered = true,
         breaklinks = true,
         linktocpage=all,
         pagebackref=false,
         colorlinks = true,
         linkcolor = BrickRed,
         urlcolor = blue,
         citecolor = BrickRed,
         anchorcolor = green,
         hyperindex = true,
         hyperfigures
         ]{hyperref}

\usepackage{cancel}
\usepackage[dvipsnames,tables]{xcolor}

\begin{document}
\begin{center}
\large{\textbf{Step by Step to Peace in Syria}}

Raphael Parens and Yaneer Bar Yam

\normalsize{{\it New England Complex Systems Institute

210 Broadway Suite 101 Cambridge MA 02139, USA}

(Dated February 9, 2016)}

\end{center}

{\bf The revolution and Civil War in Syria has led to substantial death and suffering, a massive refugee crisis, and growth of ISIS extremism and its terror attacks globally. Conflict between disparate groups is ongoing. Here we propose that interventions should be pursued to stop specific local conflicts, creating safe zones, that can be expanded gradually and serve as examples for achieving a comprehensive solution for safety, peace and stable local governance in Syria. }

\ \

Nearly five years since the deterioration of government control and order, stemming from the ``Arab Spring" protests of 2011, Syria has become a ``failed state" with widespread violence, the source of ISIS ideological violent extremism and terrorism, and the origin of massive population displacement and exodus. Often called a civil war, violence in Syria includes conflict between many parties: the government forces, ISIS, several distinct rebel groups with various affiliations, and many external parties including national actors (Saudi Arabia, Turkey, Iran, Russia and the US) and non-state actors (Hezbollah, Druze and Kurdish groups). Despite the existence of alliances on the ground, specifically in opposition to ISIS, these groups have different agendas and support different local groups and may interfere with, or even fight and kill, members of other groups. 

Because of the complexity of the conflict there are various framings of the situation as a battle between the opposition and Assad, a battle against ISIS, a battle of proxies, etc. Here we consider a strategy of achieving peace by addressing local conflicts with local solutions prior to building a comprehensive solution to the entire set of conflicts at the national scale. This is a grass roots approach to building safety and security from the ground up. Such efforts already exist in some places. We provide a framework for this approach in terms of a validated scientific analysis of essential drivers of conflict rooted in ethnic geography. While some of those involved will consider our discussion narrowly in terms of whether we provide support for their cause, our objective is to address directly the suffering of populations through establishing robust local safety and governance that will be an immediate relief to local populations severely affected by the conflict. The safety of the populace should not and need not be a hostage for the national solution which may follow. 

\begin{figure}[t]
\begin{center}
\includegraphics[width=0.68\columnwidth]{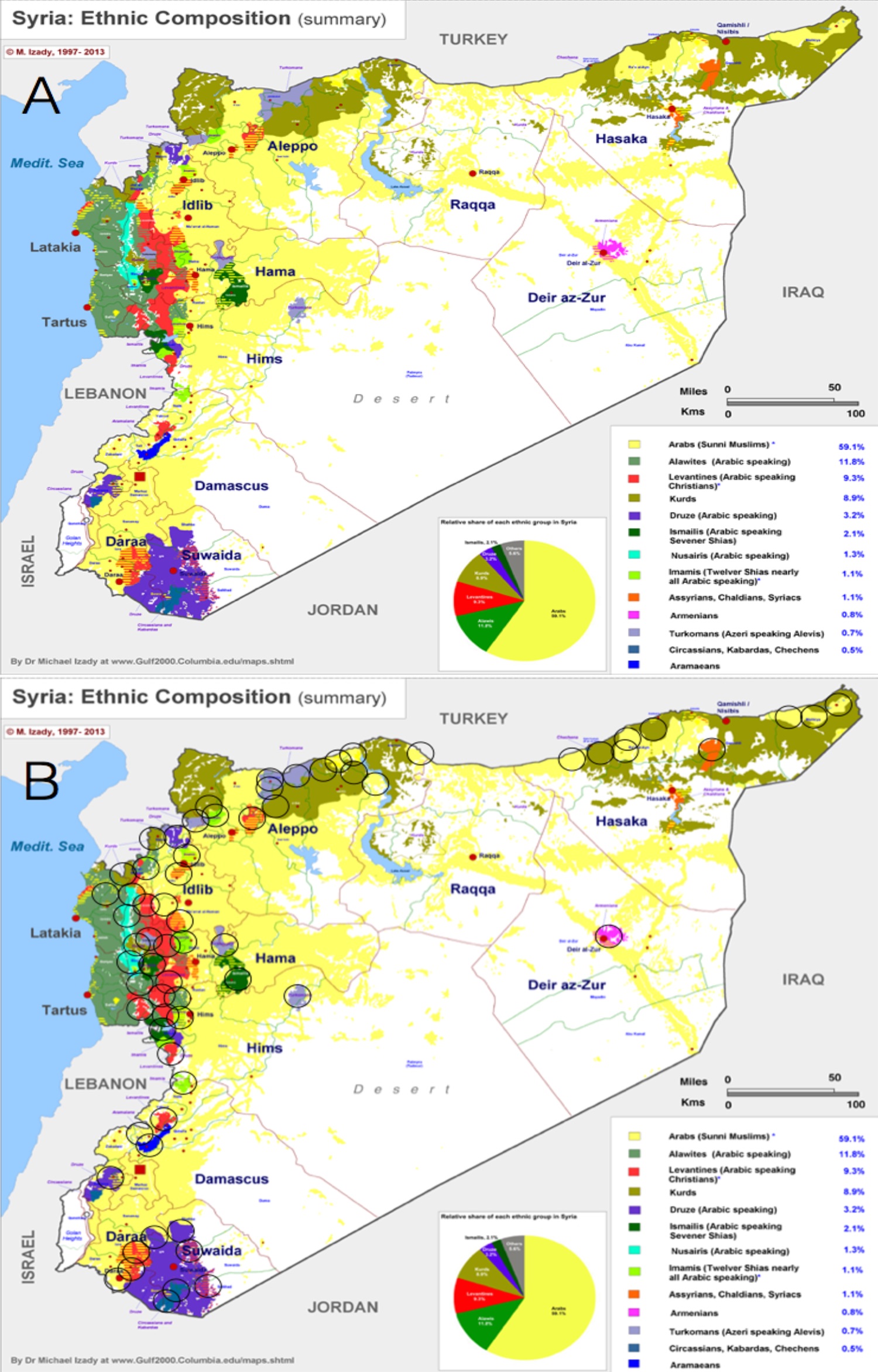}
\caption{\label{fig:Ethnicity} \textbf{Ethnic Geography of Syria} A. Ethnic geography of Syria \cite{Izady2000}. B. Superimposed circles indicate areas of a patch size indicating the likelihood of ethnic violence \cite{Lim2007,BarYam2013}.}
\end{center}
\end{figure}

Complex ethnic geography is a central reason for the large number of different groups in conflict as this geography leads to local allegiances that do not aggregate at the national scale. The makeup, and even existence of several of these groups is a matter of debate, complicated by the possibility that some groups  choose to hide their differences (e.g. Alawites and Nusaries) \cite{Lewis2014}. These debates do not affect the overall framework of our discussion, although they may play a role in the subsequent consideration of local conditions. A well researched map of ethnic geography is shown in Fig. \ref{fig:Ethnicity} showing 13 ethnic groups that have local geographic areas in which they are the majority population. The largest of these ethnic groups are, in descending order by population size, Arab Sunni Muslims (the majority, 60\% of the population), Arabic speaking Alawites, Arabic speaking Christians (Levantines), Kurds, Druze, Ismailis, Nusaries, and Imamis \cite{Izady2000}.  

We have previously demonstrated that, where ethnic groups exist in geographic patches of 20-60 km in diameter, there is a high probability of conflict \cite{Lim2007}. We have also shown that establishing local autonomy through subnational boundaries (as is found in Switzerland) is a means of alleviating conflict \cite{Rutherford2014}. Our analysis suggests that providing some level of local autonomy to the ethnic groups would reduce the impetus for local conflict and could serve as a basis for peace and stability \cite{BarYam2013}. 

The objective of local autonomy is to enable local decision making to reflect values in determining ordinances, institutions and services to the local population. As is found in Switzerland, where there are multiple cantons of each of catholic and protestant dominated areas, there is no need to have all of the members of a group in a single governance structure. The objective rather, is to reduce friction that leads to conflict by enabling those with widely different values to impose their values on public spaces. We note that the establishment of local partially autonomous regions does not itself determine the structure of federal national governance, which would be the framework for national security and interactions among the local groups as in, for example, the federal government of Switzerland. Other countries have federal governance systems with various ways of balancing local and national governance roles, including the United States, where four levels of governance are present in municipal, county, state and federal systems. 

In order to advance the understanding of the conflict and potential solutions in Syria, we have constructed a visualization identifying areas susceptible to violence. This study used a simplified methodology compared to previous studies \cite{Rutherford2014, Lim2007}. We superimposed circles of 20 km diameter on the ethnic map of Syria by visual inspection. Circles are placed where one ethnic population is the majority within the circle and others are the majority around the periphery. Fig. \ref{fig:Ethnicity} manifests why and where ethnic group affiliation, whether defined by language, religion, or both, plays a major role in the conflict in Syria. 

\begin{figure}[t]
\begin{center}
\includegraphics[scale=0.30]{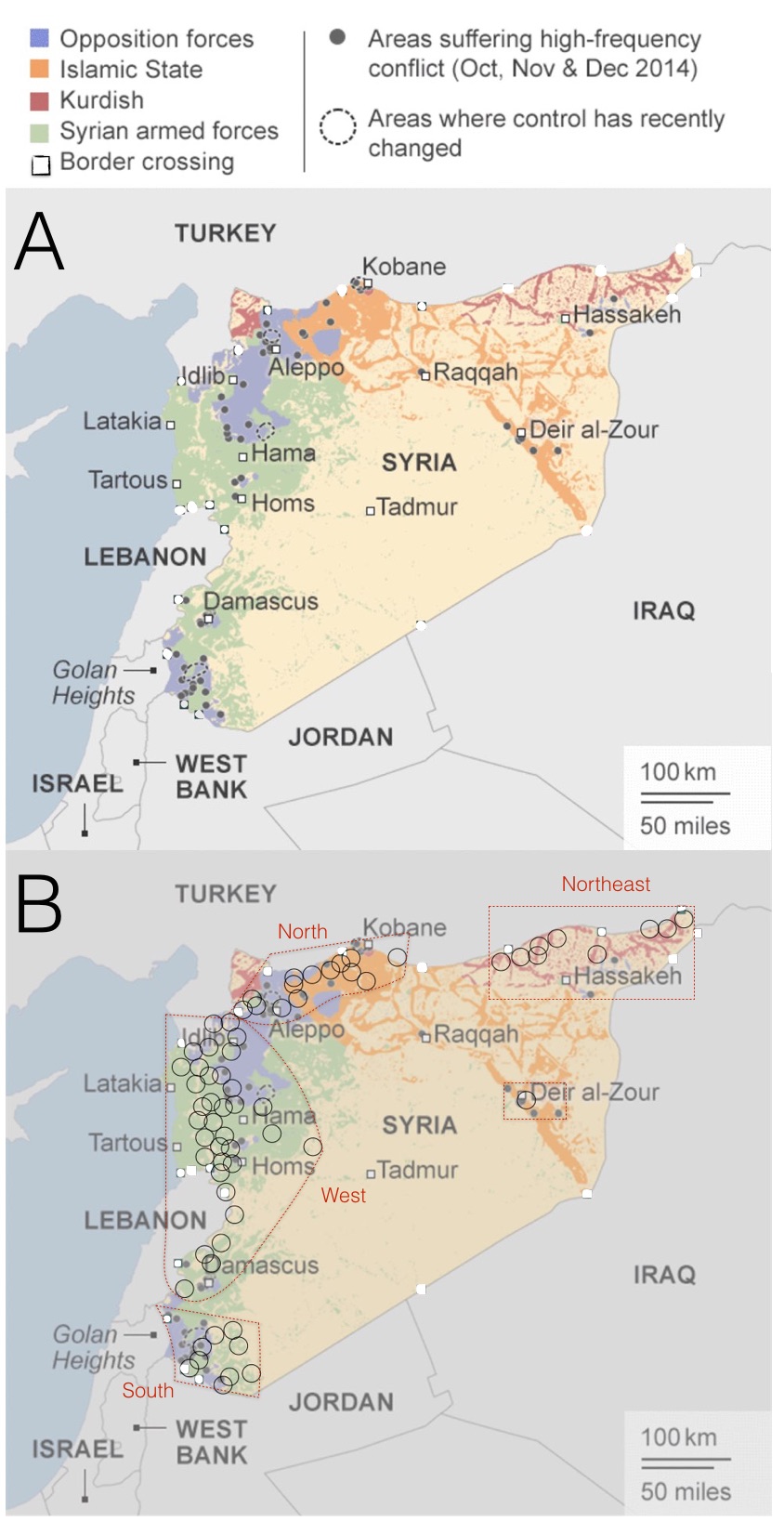}
\end{center}
\caption{\label{fig:Dubs} 
\textbf{Conflict in Syria} A. BBC map of violence in Syria, July 2015 \cite{BBC2015}. B. Superimposed locations of likely locations of violence according to our analysis.%
}
\end{figure} 

Our analytic results for the locations of likely violence are consistent with the current state of conflict shown in Fig. \ref{fig:Dubs}. We see that ethnic violence in Syria can be divided into four main regions and an isolated patch: northeast, north, west, southwest, and Deir az-Zur. 

\textit{Northeast}---The northeast region is dominated by Kurd and Sunni Arab groups. Violence is particularly likely between them along the northern border with Turkey, where the two groups are interspersed in 20 km patches. In addition, some combination of Armenians, Assyrians/Syriacs/Chaldians, and Chechens are surrounded by Kurds and Sunnis in several patches in this area, perhaps also contributing to violence. 

\textit{North}---The north region contains areas of likely conflict between Kurds, Sunnis, Circassians, Imamis, Levantines, and Turkomans. 

\textit{West}---The west region contains the largest number of likely ethnic violence areas, as a collection of Kurds, Druze, Imamis, Circassians, Assyrians/Chaldians/Syriacs, Ismailis, Nusairis, Turkomans, and Arabs all live in ethnic pockets of the critical size. 

\textit{South}---The south region in the area of Damascus also contains a significant number of ethnic patches, with possible conflict between Levantines, Druze, Aramaeans, Assyrians/Chaldians/Syriacs, Circassians, and Kabardas. 

\textit{Deir az-Zur}---Finally, the city of Deir az-Zur contains an isolated ethnic patch, where a population of Armenians is surrounded by Arab populations.  

According to our previous analysis, the establishment of boundaries between ethnic populations to provide partial local autonomy would increase stability and inhibit current and future conflict in Syria. Some suitable natural barriers already exist due to the topography of mountains, lakes, and rivers. However, many boundaries will have to be established through political borders or artificial barriers. 

Under the current conditions of fragmentary control and multiple competing groups, implementing a national process to resolve all conflicts is difficult. The complexity of local conflicts, and the many parties involved, will be a barrier to any such comprehensive plan. We propose that a step-by-step bottom up strategy provides a useful and robust alternative to a national plan. Indeed, in this context, the natural scale of intervention is at the community level. Our analysis should serve as a motivation for local governance creation and maintenance in Syria, rather than a blueprint. The complexity of governance creation on the ground will require adaptation due to the specifics of local conditions.

In such a step-by-step strategy ``safe zones'' should be established. Efforts should be made to identify specific local areas of conflict for intervention to establish the safe zones, including local governance and sub-national political borders, potentially redrawing governorates (muhafazat), districts (manatiq), or subdistricts. Such interventions should recognize specific conditions of villages and urban neighborhoods, their values and capacity to provide for their own security, as well as the relationships they have with nearby groups and national or international groups. The complexity of these local conditions must be addressed through direct engagement with the local population as an integral part of the process of achieving security. In recent years new local governance structures have emerged in response to the disorder \cite{Kirkpatrick2013, Hof2015}. These may serve as a basis for the robust and longer-term structures that are needed. Once established, a safe zone will allow the local population to rebuild and reestablish normal lives.  

Among the challenges to be faced in achieving safe zones is establishing reliable international support for security, and where necessary, assigning ownership of geographically associated economic resources and religious/cultural site control. Identifying locations where the framework for peace, even if complex, can be more readily achieved is key to early progress and establishing precedent for later more difficult areas. 

The greatest challenge in implementing this approach may be the psychological shift from seeing power as an absolute quality on a nation-state basis, and shifting to a perspective in which local power is balanced against the role of allegiances and against larger scales of power, including national. The current conflict is often seen as an irreconcilable power struggle between local groups and the nation-state, as embodied by its government. Here the objective is to show how local groups can coexist with national power. This coexistence is not about full autonomy but rather about a balance between local autonomy within relationships among the local, national and international groups. 

By focusing on the local nature of interactions, the basis of community governance can be established. The structure of national power, in whatever form then occurs, becomes secondary as individual day-to-day existence can be primarily determined by the local authority and only secondarily with the national governance.

\clearpage
\textbf{Acknowledgements: }

We would like to thank Francisco Prieto-Castrillo, Nima Dehghani and Matthew Hardcastle for helpful comments on the manuscript and Nadim Shehadi, Nassim Taleb and Tom Garvey for helpful discussions.

\end{document}